\documentclass[a4,12pt,epsfig,graphicx]{article}
\usepackage{pstricks}
\usepackage{graphicx}
\usepackage[latin2]{inputenc}
\usepackage{epic}
\usepackage{latexsym}
\usepackage{amssymb}
\def\beq{\begin{equation}} 
\def\eeq{\end{equation}} 
\def\bea{\begin{eqnarray}}  
\def\eea{\end{eqnarray}}  
\def\bq{\begin{quote}}  
\def\eq{\end{quote}}  
\def\bi{\begin{itemize}}  
\def\ei{\end{itemize}}  

\def\be{\begin{enumerate}}  
\def\ee{\end{enumerate}}  

\def\bc{\begin{center}}
\def\ec{\end{center}}  

\def\pa{\partial}

\def\pr{\otimes}

\def\ta{\tilde a}

\def\cl{{\cal L}}

\def\r2{\sqrt{2}} \def\rt{\sqrt{2}} 
\def\ra{\rightarrow}

\def\ov{\overline}  
\def\nn{\nonumber \\}

\newsavebox{\cmoose}
\sbox{\cmoose}{%
\begin{picture}(0,0)
  \thicklines
  \put(-60,0){\circle{35}}
  \put(60,0){\circle{35}}
  \dashline{10}(-40,0)(40,0)
 \put(0,0){\vector(1,0){0}} 
\end{picture}}
\newcommand{\NPB}[3]{\emph{ Nucl.~Phys.} \textbf{B#1} (#2) #3}   
   
\newcommand{\PRD}[3]{\emph{ Phys.~Rev.} \textbf{D#1} (#2) #3}   
\newcommand{\PRL}[3]{\emph{ Phys.~Rev.~Lett.} \textbf{#1} (#2) #3}

\newcommand{\JHEP}[3]{\emph{JHEP} \textbf{#1} (#2) #3}

\begin{document}
\pagestyle{empty}
\setcounter{page}{0}
{\normalsize\sf
\rightline {hep-th/0204195}
\rightline{IFT-02/14}
\rightline{Saclay T/02-010}
\vskip 3mm
\rm\rightline{April 2002}
}
\vskip 2cm 
\begin{center} 
{\Huge Custodial supersymmetry\\ in non-supersymmetric quiver theories
}
\vspace*{5mm} \vspace*{1cm}  
\end{center} 
\vspace*{5mm} \noindent 
\vskip 0.5cm 
\centerline{\bf Philippe Brax${}^{1}$, Adam Falkowski${}^{2}$, Zygmunt Lalak${}^{2}$ {\rm and} Stefan Pokorski${}^{2}$} 
\vskip 1cm 
\centerline{\em ${}^{1}$ Service de Physique Th\'eorique} 
\centerline{\em CEA-Saclay F-91191 Gif/Yvette, France} 
\vskip 0.3cm 
\centerline{\em ${}^{2}$Institute of Theoretical Physics} 
\centerline{\em University of Warsaw, Poland} 
\vskip 2cm 
 
\centerline{\bf Abstract} 
\vskip 0.5cm
\noindent We consider non-supersymmetric quiver  theories obtained by orbifolding the ${\cal N}=4$ supersymmetric $U(K)$ gauge theory by a discrete $Z_\Gamma$ group embedded in the $SU(4)$ R-symmetry group. We explicitly find that in such theories there are no one-loop quadratic divergences in the effective potential. Moreover, when the gauge group $U(n)^\Gamma$ of the  quiver theory  is spontaneously broken down to the diagonal $U(n)$, we identify a custodial supersymmetry which is responsible for the fermion-boson degeneracy of the mass spectrum.

\vskip .3cm 
\newpage

\setcounter{page}{1} \pagestyle{plain}
A class of interesting four-dimensional gauge theories has been studied for some time in the context of AdS/CFT conjecture and, more generally, in the search for 4d (super)conformal theories.  These are so called quiver theories \cite{DOMO,GR,KASI} constructed by studying 
$K=\sum_i k_i$ $D3$ branes on orbifolds of ${\cal R}^6$ or, equivalently, by orbifolding the ${\cal N}=1$ $U(K)$ gauge theory in 10d reduced down to 4d.  They can have ${\cal N}=4$, ${\cal N}=2$,   ${\cal N}=1$ or ${\cal N}=0$ supersymmetry. Such theories have an equal number of fermions and bosons but in the non-supersymmetric theories, which we discuss in this paper, fermions and bosons transform according to different representations of the gauge group. 

Apart from purely theoretical interest one may ask about a phenomenological relevance of such theories. Actually, similar gauge  theories have been recently discussed for  different reasons.  It has been observed \cite{ARCOGE_dec,HIPOWA} that renormalizable gauge theories in four dimensions with the gauge group $U(n)^\Gamma$ (or $SU(n)^\Gamma$) and scalars 
in bifundamental representations appear to be equivalent to a higher dimensional $U(n)$ (or $SU(n)$) theory  below some deconstruction scale $v$.
 This energy scale  is set by the expectation values of the scalar fields that break the group $U(n)^\Gamma$ to its diagonal subgroup. These setups put new 
twists on  gauge coupling unification \cite{ARCOGE_acc}, supersymmetry breaking \cite{ARCOGE_twi}, the hierarchy problem \cite{ARCOGE_ele} and other issues \cite{SKSM}.  

Thus, the orbifold constructions provide a possible derivation of, at least, some of the models of deconstructed dimensions\footnote{Some relations between quiver and deconstruction theories, in the large $n$ limit, were discussed in \cite{ROSK}}. The question then arises about the origin of the diagonal breaking, the origin of supersymmetry breaking, stability of the corresponding scales, etc. The models with ${\cal N}=2$ and   ${\cal N}=1$ are superconformal (if the $U(1)$ factors are decoupled) \cite{LANEVA,FRVA}. Their finiteness is an interesting issue in its own right because hierarchies of scales are automatically protected, however generating a specific value of  the diagonal breaking or supersymmetry breaking would require some additional physics.

 Non-supersymmetric quiver theories  are not conformal for finite $n$ \cite{CSSKTE}, but precisely because of that they may be phenomenologically more interesting.  
These theories are not finite, therefore in principle they offer the possibility of calculating  vevs of  scalars in terms of the cut-off scale $\Lambda$. 
Independently, once a hierarchically small vev, $v/\Lambda\ll 1$, is introduced, one wishes this hierarchy not to be destroyed by loop corrections.
In this paper we discuss such quiver theories and point out two interesting properties, of relevance for phenomenology.  
We show that indeed the one-loop effective potential of non-supersymmetric quiver theories has, under some mild assumptions, no quadratic divergences ($STr M^2=0$). Thus the minimization of the effective potential would be quadratically sensitive to the scale $\Lambda$ not earlier than 
at two-loops, and the hierachy $v/\Lambda\ll 1$ would be protected at one-loop.
Note that implicit in these considerations is  the idea that the cut-off scale $\Lambda$ may be much lower than the Planck scale. Otherwise,
for very large $\Lambda$ the quadratic sensitivity to unknown physics at 
higher loops would 
reintroduce  the usual hierarchy problem. Similar point of view on the hierarchy problem has been recently  taken in a number of papers, e.g. in \cite{CRIBMA,FAGRPO}.  

Furthermore, we discuss models assuming the diagonal symmetry  breaking. There are two basic points to be made here. First, the massless sector exhibits ${\cal N}=4$ supersymmetry. Second, we show that the model  in the deconstruction phase has a custodial supersymmetry, which is responsible for the fermion-boson degeneracy of the massive spectrum at the tree-level.  As a  consequence of custodial supersymmetry the massless modes remain massless also at one-loop. However, since the theory is non-supersymmetric, we expect the mass splittings to appear at the two-loop level. Another consequence of the custodial supersymmetry is that, for universal vevs, 
$Str M^{2q} =0$ for any $q$, thus universal vevs remain flat 
directions of the effective potential at one loop. 
In short, we show that the class of spontaneosly broken quiver theories 
discussed here may be phenomenologically interesting, as it supports the 
phenomenologically relevant hierarchy of scales: 
$M_{SUSY} \ll v \ll \Lambda$.  

The conventions we use throughout this paper are the following. The indices  index $i$,$j$,$k$ take values $1 \dots 3$ and label the three generations of chiral multiplets in ${\cal N}=4$ supersymmetric lagrangian. The indices $p$,$r$ take values $1 \dots \Gamma$ and label the consecutive $n \times n$ blocks within the $K \times K$ matrices, or, equivalently count the $U(n)$ group in the  $U(n)^\Gamma$ product group. Finally, $(n) = 0 \dots (\Gamma-1)/2$  labels the mass level in the deconstruction phase of the theory. 

Consider a four-dimensional ${\cal N}=4$ supersymmetric  mother theory with  the gauge group $U(K)$. Its field content is the gauge field $A$, four fermions $\psi_i$, $\psi_4 \equiv \lambda$  and three scalars $\phi_i$, all in adjoint representation of $U(K)$. The lagrangian is:
\bea 
\label{eq:n4lagrangian}
&\cl= 
Tr \left \{
-{1 \over 2} F_{\mu \nu} F_{\mu\nu} + i \ov{\lambda} \gamma^\mu D_\mu \lambda  
+ 2 D_\mu \phi_i^\dagger D_\mu \phi_i  + i \ov{\psi_i} \gamma^\mu D_\mu \psi_i
\right. &\nn&
- 2i\rt g_0 ( \ov{\psi_i} [P_R \lambda,\phi_i] +  \ov{\psi_i} [P_L \lambda,\phi_i^\dagger] )
-i\rt g_0 \epsilon_{ijk} (\ov{\psi_i} [P_L \psi_j,\phi_k] +
\ov{\psi_i} [P_R \psi_j,\phi_k^\dagger])
 &\nn& \left .
 - g_0^2[\phi_i, \phi_i^\dagger] [\phi_j, \phi_j^\dagger] 
+2  g_0^2 [\phi_i, \phi_j][\phi_i^\dagger, \phi_j^\dagger]
\right \}.
\eea    
All fermions satisfy the Majorana condition. 
By introducing 
$\phi_{ij} = \epsilon_{ijk} \phi_{k}$ the $SU(4)$ R-symmetry becomes explicit, but we do not make use of such notation in the following. 

 One can obtain a daughter theory with fewer supersymmetries than  ${\cal N}=4$ by means of orbifolding \footnote{ For a transparent review of the orbifolding procedure see \cite{SC}.}, that is dividing the mother theory by a discrete group embedded simultaneously in  the gauge group  and in the R-symmetry group. One  retains in the spectrum only those fields which  are invariant under the action of the dicrete group. The interactions of the daughter theory are inherited from the mother theory, with all non-invariant fields and their 
interactions removed . 

If the discrete group is  $Z_\Gamma$, the matrix $\gamma$, which represents the embedding of $Z_\Gamma$  into $U(K)$,  can be chosen as a direct sum of $\Gamma$ unit matrices of dimensions $k_x \times k_x$, $x=0,...,\Gamma-1$, $\sum k_x = K$, each multiplied respectively by $\omega^x$ with $\omega=e^{ \frac{2 \pi}{\Gamma} i}$. 
The invariant components of the gauge fields fulfill the condition 
\beq
A = \gamma A \gamma^{-1}.
\eeq
This leaves invariant the product group $\prod_{i=x}^{\Gamma} U(k_x)$ as only diagonal $k_x \times k_x $ blocks are left  invariant.

The invariant components of fermions and scalars satisfy the condition:   
\bea
\psi_i &=& \omega^{a_i} \gamma \psi_i \gamma^{-1}, \;\;\;i=1..4,
\nn
\phi_i &=& \omega^{\tilde{a}_j} \gamma \phi_i \gamma^{-1}, \;\;\;i=1..3, 
\eea
where the integers (shifts) $a_i$, $\ta_i$ obey the constraints: 
\begin{equation}
a_1+a_2 +a_3 + a_4 =0, \hspace{2cm} \tilde a_i=a_i+a_4. 
\end{equation} 

In this paper we concentrate on the  situation  when  $k_x=n$, $n \, \Gamma = K$, i.e.  $\gamma$ consists of $n$ copies of  the  regular representation of $Z_\Gamma$. In such a  case the daughter theory gauge group consists of  $\Gamma$ copies of $U(n)$ (denoted $U(n)^\Gamma$). It is convenient to divide the mother theory fields into $\Gamma \times \Gamma$ equal size square blocks.  
The truncated fields have the  block structure:
\bea
\label{eq:bd}
A_{pr} &=& A_p \delta_{p,r}
\nn
\phi^i_{pr} &=&\phi_{i,p}\delta_{p,r-\tilde a_i}
\nn
\psi_{pr}^i&=&\psi_{i,p}\delta_{p,r-a_i}.
\eea
Obviously the gauge field $A_p$  transforms in adjoint representation $\bf ( n_p,\ov{n}_{p})$, while the fermion $\psi_{i,p}$ and scalar $\phi_{i,p}$ fields  transform respectively as $\bf ( n_p,\ov{n}_{p+a_i})$,  $\bf ( n_p,\ov{n}_{p+\ta_i})$ representations of the $(U(n))^\Gamma$ group.    

An easy way to count unbroken supersymmetries in the daughter theory  is to analyse  the number of  fermions in  the adjoint representation in each factor of the gauge group, i.e. to count the gauginos. A straightforward observation is that a situation where some factors of the gauge group have 
invariant gauginos while others do not  is excluded by the construction.
Supersymmetry is preserved when  the group $Z_{\Gamma}$ is embedded in $SU(3)$ ($a_i=0$ or $a_4=0$), when  at least one of the fermionic representation is the adjoint representation and i.e. when gauginos of ${\cal N}=1$ supersymmetry are present. Embedding the orbifold group in $SU(2)$  leads to the presence of two gauginos and ${\cal N}=2$ supersymmetry. 

As we explain in more details at the end of this paper, such particular construction is motivated by string theory considerations. More precisely, the mother theory  describes the low-energy action for  a  stack of $n$  coinciding D3 branes in the type IIB string theory. The orbifolding down to the daughter theory corresponds to orbifolding  the six-dimensional space transverse to the D-branes. 

Our first task is to determine the $UV$ properties of the one-loop effective potential of the daughter theory. To this end we need to calculate the mass 
matrices of gauge bosons, scalars and fermions, with the expectation value of the scalar fields switched-on.  
We will show that, under certain conditions, the supertrace of  ${\cal M}^2$ vanishes, hence the  one-loop  quadratic divergences are absent in the daughter theory. 

Inserting the block decomposition (\ref{eq:bd}) into the ${\cal N} =4$ lagrangian we find the daughter theory lagrangian: 
\bea &
\label{eq:daughter}
\cl= 
Tr \left \{
-{1 \over 2} F_{\mu \nu,p} F_{\mu\nu,p} + i \ov{\lambda_p} \gamma^\mu D_\mu \lambda_p 
+ 2 D_\mu \phi_{i,p}^\dagger D_\mu \phi_{i,p}  + i \ov{\psi_{i,p}} \gamma^\mu D_\mu \psi_{i,p} \right .
&\nn&
- g_0\left [2i\rt ( \ov{\psi_{i,p}} P_L \lambda_{p+a_i} \phi_{i,p}^\dagger -  \ov{\psi_i}  \phi_{i,p-a_4}^\dagger P_L \lambda_{p-a_4})
+ {\rm h.c.} \right]
&\nn&
-g_0\left [i\rt \epsilon_{ijk} (
\ov{\psi_{i,p}} P_L \psi_{j,p+a_i} \phi_{k,p-\tilde a_k}
 - \ov{\psi_{i,p}} \phi_{k,p+a_i} P_L \psi_{j,p-a_j} ) + {\rm h.c.}
\right ]
&\nn& 
-g_0^2(\phi_{i,p} \phi_{i,p}^\dagger 
- \phi_{i,p-\tilde a_i}^\dagger\phi_{i,p -\tilde a_i }) 
(\phi_{j,p} \phi_{j,p}^\dagger 
- \phi_{j,p-\tilde a_j}^\dagger\phi_{j,p -\tilde a_j })
&\nn& \left.
+4 g_0^2 (\phi_{i,p} \phi_{j,p+ \tilde{a}_i} 
 \phi_{i,p+ \tilde{a}_j}^\dagger  \phi_{j,p}^\dagger
- \phi_{i,p} \phi_{j,p+ \tilde{a}_i} \phi_{j,p+ \tilde{a}_i}^\dagger \phi_{i,p}^\dagger) \right \}.
&\nn&
\eea
The covariant derivative acting on scalars is 
$D_\mu\phi_{i,p} = \pa_\mu \phi_{i,p} + ig_0A_p\phi_{i,p} -ig_0\phi_{i,p}A_{p+\ta_i}$. Hence  the squared mass matrix of the gauge bosons is :
\beq 
({\cal M}_A^2)_{pr} = 2 g_0^2 \left[ 
 \delta_{p,r} 1 \otimes \phi_{k,p} \phi_{k,p}^\dagger 
+\delta_{p,r} \phi_{k,p-\ta_k}^\dagger  \phi_{k,p - \ta_k} \otimes 1
-\delta_{p+\ta_k,r} \phi_{k,p} \otimes \phi_{k,p}^\dagger 
-\delta_ {p,r+\ta_k} \phi_{k,p-\ta_k}^\dagger \otimes \phi_{k,p-\ta_k} \right].
\eeq
The tensor product defines the way the scalars  are contracted with the gauge fields. The rule is $A_1 (X \otimes Y) A_2 = A_1 X A_2 Y$.

In order to extract the scalar mass terms we substitute 
$\phi_{i,p} \ra \phi_{i,p} + X_{i,p}$ in the scalar potential and calculate the terms which are quadratic in the fluctuations  $X$. Moreover, it is convenient to separate the F-term and the D-term  contributions. 
We define the mass matrix of scalars as:
\beq 
\cl = -2 (\begin{array}{c} X_{i,p} \\ X^\dagger_{i,p} \end{array})^\dagger 
\left ( \begin{array}{cc}
{1\over 2} ({\cal M}_F^2 +{\cal M}_D^2 )^{ij}_{pr} &(M_F^2+M_D^2)^{ij}_{pr} \\
(M_F^{2\dagger}+ M_D^{2\dagger})^{ij}_{pr} & {1\over 2} ({\cal M}_F^2 +{\cal M}_D^2 )^{ij}_{pr}
\end{array} \right)
(\begin{array}{c} X_{j,r} \\ X^\dagger_{j,r} \end{array}).
\eeq

The diagonal entries are: 
\bea 
({\cal M}_F^2 )^{ij}_{pr} = &
2 g_0^2 \delta_{ij} [
\delta_{p,r} \phi_{k,p-\ta_k}^\dagger  \phi_{k,p-\ta_k} \pr 1 
+\delta_{p,r} 1\pr \phi_{k,p + \ta_i} \phi_{k,p + \ta_i}^\dagger
&\nn& -\delta_{p,r+\ta_k}\phi_{k,p-\ta_k}^\dagger \pr \phi_{k,p +\ta_i -\ta_k} 
- \delta_{p,r-\ta_k}\phi_{k,p}\pr \phi_{k,p+\ta_i}^\dagger
]
&\nn& +2 g_0^2 [
- \delta_{r,p +\ta_i-\ta_j} \phi_{j,p-\ta_j}^\dagger  \phi_{i,p-\ta_j} \pr 1 
-\delta_{p,r} 1 \pr \phi_{i,p+\ta_j} \phi_{j,p+\ta_i}^\dagger
&\nn&
+\delta_{p,r-\ta_i}\phi_{i,p} \pr \phi_{j,p +\ta_i }^\dagger
+\delta_{p,r+\ta_j}\phi_{j,p -\ta_j}^\dagger \pr \phi_{i,p} ], 
\eea

\bea 
&({\cal M}_D^2 )^{ij}_{pr} = 
g_0^2 \delta_{ij}\delta_{p,r}[
\phi_{k,p} \phi_{k,p}^\dagger\pr 1
+ 1 \pr \phi_{k,p + \ta_i-\tilde a_k}^\dagger  \phi_{k,p + \ta_i-\tilde a_k}
-\phi_{k,p-\tilde a_k }^\dagger \phi_{k,p -\tilde a_k} \pr 1 
-1 \pr \phi_{k,p + \ta_i}  \phi_{k,p + \ta_i}^\dagger]
&\nn&  
+g_0^2[\delta_{r,p +\ta_i-\ta_j }\phi_{i,p}  \phi_{j,p +\ta_i-\ta_j }^\dagger \pr 1 
+ \delta_{p,r} 1 \pr \phi_{j,p}^\dagger  \phi_{i,p}
-\delta_{p,r-\ta_i}\phi_{i,p} \pr \phi_{j,p +\ta_i }^\dagger
- \delta_{p,r+\ta_j} \phi_{j,p -\ta_j}^\dagger \pr \phi_{i,p} ]. 
\eea

Finally, the squares of the fermion mass matrices are:
\bea
({\cal M}_\Psi^\dagger)^{ik}_{ps}({\cal M}_\Psi)^{kj}_{sr}+
({\cal M}_\Psi^\dagger)^{i4}_{ps}({\cal M}_\Psi)^{4j}_{sr}  = 
 &
2 g_0^2 \delta_{ij} [
\delta_{p,r} \phi_{k,p-\ta_k}^\dagger  \phi_{k,p-\ta_k} \pr 1 
+\delta_{p,r} 1\pr \phi_{k,p + a_i} \phi_{k,p + a_i}^\dagger
&\nn& -\delta_{p,r+\ta_k}\phi_{k,p-\ta_k}^\dagger \pr \phi_{k,p +a_i -\ta_k} 
- \delta_{p,r-\ta_k}\phi_{k,p}\pr \phi_{k,p+a_i}^\dagger
]
&\nn& +2g_0^2 [
- \delta_{p,r +\ta_j-\ta_i} \phi_{j,p-\ta_j}^\dagger  \phi_{i,p-\ta_j} \pr 1 
-\delta_{p,r} 1 \pr \phi_{i,p+a_j} \phi_{j,p+a_i}^\dagger
&\nn&
 \delta_{p,r +\ta_j-\ta_i }\phi_{i,p}  \phi_{j,p +\ta_i-\ta_j }^\dagger \pr 1 
+ \delta_{p,r} 1 \pr \phi_{j,p-a_4}^\dagger  \phi_{i,p-a_4}], 
\eea
\bea
({\cal M}_\Psi^\dagger)^{4k}_{ps}({\cal M}_\Psi)^{k4}_{sr} = 
 &2 g_0^2[
 \delta_{p,r}  \phi_{k,p} \phi_{k,p}^\dagger \pr 1 
+\delta_{p,r} 1 \pr \phi_{k,p- a_k}^\dagger  \phi_{k,p - a_k} 
&\nn&
-\delta_{p,r-\ta_k} \phi_{k,p} \otimes \phi_{k,p+a_4}^\dagger 
-\delta_ {p,r+\ta_k} \phi_{k,p-\ta_k}^\dagger \otimes \phi_{k,p-a_k}].
\eea

Taking into account the number of degrees of freedom and statistics of each field, the supertrace is:
\beq
STr ({\cal M}^2) = 3 Tr ({\cal M}_A^2) + 2  Tr ({\cal M}_F^2)
+2 Tr ({\cal M}_D^2) - 2 Tr ({\cal M}_\Psi^\dagger {\cal M}_\Psi)^{44}
-2 Tr ({\cal M}_\Psi^\dagger {\cal M}_\Psi)^{ij}.
\eeq

Finally we find the supertrace of the mass matrix: 
\bea
\label{eq:str}
STr ({\cal M}^2) =
4g_0^2 \sum_k \sum_p \delta_{\ta_k,0} \left [   
\left ( Tr (\phi_{k,p}^\dagger) Tr(\phi_{k,p+a_4})
+Tr(\phi_{k,p+a_4}^\dagger)Tr (\phi_{k,p})
-2 Tr (\phi_{k,p}^\dagger) Tr(\phi_{k,p}) \right)\right.
\nn + \left . \sum_i
 \left(
Tr (\phi_{k,p}^\dagger) Tr(\phi_{k,p+a_i}) 
+  Tr(\phi_{k,p+a_i} ^\dagger)Tr (\phi_{k,p}) 
-Tr (\phi_{k,p}^\dagger) Tr(\phi_{k,p+\ta_i}) 
-  Tr(\phi_{k,p+\ta_i} ^\dagger)Tr (\phi_{k,p}) \right )
\right].
&\nn&
\eea

One can check  that (\ref{eq:str}) vanishes identically if at least one of the following conditions is satisfied:
\bi
\item $a_4=0$ or $a_i=0$, that is when at least ${\cal N=1}$ supersymmetry is preserved by the orbifolding,  
\item $\ta_1 \neq 0 $, $\ta_2 \neq 0$  $\ta_3 \neq 0$, that is when there are no scalars in adjoint representation of $U(n)$ group.
\ei 
In the first case the vanishing of the supertrace is of course guaranteed by unbroken supersymmetry of the daughter theory. Surprisingly, the absence of quadratic divergences can also occur if the daughter theory is  completely non-supersymmetric, the only condition being that all scalars are in bifundemental representations of the $U(n)^\Gamma$ gauge group. This condition is non-trivial, as for instance the choice  $\tilde a_1=\tilde a_2=\tilde a_3=0$ and $\Gamma =2$
 leads to a non-supersymmetric theory with a non-vanishing supertrace, unless we demand $Tr[\Phi_{i,p}]=0$. Note that in the non-supersymmetric case the cancellation would not occur for finite $n$ if the gauge group were $SU(n)^\Gamma$. In such case, as shown in ref. \cite{CSSKTE}, the  quadratic divervences do appear (suppressed by ${1 \over n}$) even in absence of adjoint scalars. 
One can also  show that the vanishing of the supertrace is particular to the regular representation, as only in such case the equal number of bosons and fermions in the theory is ensured.

 It turns out that `supersymmetric' properties are enhanced in the deconstruction phase, by which we mean the situation when scalar fields acquire vacuum expectation values proportional to the identity matrix and  independent of the block indices:
\beq
\label{eq:vev}
\langle \phi_{i,p} \rangle = v_i {\bf 1}_{n \times n}.
\eeq
For simplicity we assume that $v_i$'s are real. Note that we still allow the vevs to depend on the generation index.

In general, such vevs as in (\ref{eq:vev}) imply spontaneous breaking of the $U(n)^\Gamma$ down to the diagonal $U(n)$ group\footnote{However, if all $\ta_k$ are divisors of $\Gamma$ then the gauge group can be  broken  to some product of $U(n)$ groups.}. 
 The gauge bosons acquire mass terms:
\beq
\cl =  \sum_p \sum_{k=1}^3 g_0^2 v_k^2 (A_p^a - A_{p+\ta_k}^a)^2.  
\eeq
(We have rewritten the gauge fields as $A = A^aT^a$ and evaluated the trace over generators. In the following we often omit the adjoint index $a$.) These mass terms are diagonalized by the folowing mode decomposition \footnote{The 
 decomposition is given for odd $\Gamma$. For even $\Gamma$ the first sum goes to $\Gamma/2$ and the second to $\Gamma/2-1$.}:
\beq
\label{eq:gfmd}
A_p = \sqrt{2 \over \Gamma} \left ( 
\sum_{n=0}^{(\Gamma-1)/2} \eta_n \cos \left( {2 n \pi \over \Gamma}p\right)A^{(n)}
+\sum_{n=1}^{(\Gamma-1)/2}\sin \left( {2 n \pi \over \Gamma}p\right)\tilde A^{(n)} \right).
\eeq
where $\eta_0=1/\sqrt 2$ and $\eta_n=1,\ n\ne 0$.
Plugging in this decomposition we get:
\beq \label{spectr}
\cl =  {1\over 2}\sum_n\sum_k (m_k^{(n)})^2 (A^{(n)}A^{(n)} +\tilde A^{(n)}\tilde A^{(n)})\hspace{2cm} m_k^{(n)} \equiv 2\rt g_0 v_k \sin \left( {n \pi \over \Gamma}\ta_k\right),   
\eeq
so that the $n$-th level gauge bosons have masses  $(m^{(n)})^2 = \sum_k m_k^2$. 

Scalar vevs also generate masses of the fermion and scalar fields. We decompose  the complex scalars into real ones as $X_{i,p} = {1 \over \rt}(B_{i,p}^a + i C_{i,p}^a) T^a$. Then we use  the  mode decomposition (\ref{eq:gfmd}), with $p \ra p+ \ta_i/2$ on the r.h.s and we get :    
\beq
\cl_B =  -{1\over 2}\sum_i\sum_n(m^{(n)})^2 (B_i^{(n)}B_i^{(n)} +\tilde B_i^{(n)}\tilde B_i^{(n)}),  
\eeq 
\beq
\cl_C =  -{1\over 2}\sum_i\sum_j\sum_n ( 
\delta_{ij}(m^{(n)})^2 -m_i^{(n)}m_j^{(n)}) (C_i^{(n)}C_j^{(n)} +\tilde C_i^{(n)}\tilde C_j^{(n)}).  
\eeq 
 The n-th level $B$-scalar is degenerate in mass with the n-th level gauge boson. As for the $C$-scalar, the mass matrix must be further diagonalized in generation indices. This mass matrix has $\Gamma$ zero eigenvalues corresponding to eigevectors $ C^{(n)}_0 = {1\over m^{(n)}} \sum_k m_k C_k^{(n)} $, $
\tilde C^{(n)}_0 = {1\over m^{(n)}} \sum_k m_k \tilde C_k^{(n)}$. 
These eigenvectors are readily identified with the  Goldstone bosons of the spontanously broken $U(n)^\Gamma$ symmetry. All but one of them are  eaten by massive vector fields. 
The other two combinations of $C_i^{(n)}$ orthogonal to  $C_0^{(n)}$ 
acquire a mass  equal to $ m^{(n)}$.  

Finally we investigate the fermion masses in the deconstruction phase. 
The mode decomposition is again analogous to (\ref{eq:gfmd}) with $p \ra p+a_4$ for gauginos and $p \ra p+\ta_i/2$ for  $\psi_i$. 
The mass terms take the form:
 \beq
\cl=  
- \sum_n \sum_k m_k^{(n)}( \ov{\tilde \psi_{k}^{(n)}} \lambda^{(n)}
+\ov{\psi_{k}^{(n)}} \tilde \lambda^{(n)} )
-\rt \epsilon_{ijk}
\ov{\psi_{i}^{(n)}} \tilde \psi_{j}^{(n)} m_k^{(n)}. 
\eeq
Similarly to the  C-scalars we can define $
 \psi_0^{(n)} = {1\over m^{(n)}} \sum_k m_k^{(n)}\psi_{k}^{(n)}$,  $
 \tilde \psi_0^{(n)} = {1\over m^{(n)}} \sum_k m_k^{(n)}\tilde \psi_{k}^{(n)}$
such that $\psi_0^{(n)}$ ( $\tilde\psi_0^{(n)}$)  combines with $\lambda^{(n)}$ ($\tilde \lambda^{(n)}$)  to form a Dirac fermion of mass $m^{(n)}$. The two remaining linear combinations of $\psi_i^{(n)}$, which are orthogonal to $\psi_0^{(n)}$, combine with $\tilde \psi_i^{(n)}$ to form Dirac fermions of mass $m^{(n)}$.

We see that  at the $n$-th level the spectrum is perfectly boson-fermion degenetate. This suggests that a part of the lagrangian in the deconstruction phase may possess some boson-fermion symmetry. Indeed, this is the case, 
and we shall call that approximate symmetry a custodial supersymmetry. 
A way to describe it  is to  rewrite the deconstruction phase lagrangian using the superspace formalism. We define the  vector superfields in the Wess-Zumino gauge as:
\beq
V^{(n)}(y,\theta) = {i \over 2}(\ov{\theta} \gamma_5 \gamma_\mu \theta) A^{(n)}
  -i  (\ov{\theta} \gamma_5 \theta) (\ov{\theta} \lambda^{(n)})
-{1 \over 4} (\ov{\theta} \gamma_5 \theta)^2 D^{(n)}.
\label{t1}
\eeq
Similarly we define chiral superfields:
\beq
\Phi_i^{(n)}(y,\theta) = X_i^{(n)} - \rt  (\ov{\theta} P_L \psi_i^{(n)})
+ F_i^{(n)}(\ov{\theta} P_L \theta).
\label{t2}
\eeq
Analogous expressions for the tilded fields hold. 

First, we note that the self-couplings in  the zero-mode sector  are those of the ${\cal N}=4$ supersymmetric theory. Indeed,  the interactions of the zero-modes can be found by making in (\ref{eq:daughter})  the replacement $\phi_{i,p} \ra {1 \over \sqrt{\Gamma}} \phi_i^{(0)}$ (and similarly for fermion and gauge fields). Since all memory of the block indices is lost,
as a result we obtain  the lagrangian (\ref{eq:n4lagrangian})  with the gauge coupling $g = {g_0 \over \sqrt{\Gamma}}$. Second, we have already shown that the mass pattern in the deconstruction phase is supersymmetric. It turns out that the custodial supersymmetry has a much wider extent and all the terms quadratic in the heavy modes (including triple and quartic interactions with the zero-modes) match the structure of a globally supersymmetric theory! As an example we present a superfield lagrangian which reproduces the  Yukawa terms and the scalar potential of the daughter theory:    
\bea
&\cl = \sum_n \sum_k {\rm Tr} \left [ 
4 g_0 v_k \sin \left({ n \pi \ta_k \over \Gamma}\right) \left (
 \tilde V^{(n)} \Phi_k^{(n)} -   V^{(n)} \tilde \Phi_k^{(n)} \right)
\right. &\nn&   
+ 2 g  \cos \left({ n \pi \ta_k \over \Gamma}\right)
 \left ( [\Phi_k^{(0)\dagger},\Phi_k^{(n)}] V^{(n)}  +  
[\Phi_k^{(0)\dagger},\tilde \Phi_k^{(n)}] \tilde V^{(n)} \right)
&\nn& \left . 
+  2 g  \sin \left({ n \pi \ta_k \over \Gamma}\right)
 \left ( \{ \Phi_k^{(0)\dagger},\Phi_k^{(n)} \} \tilde V^{(n)}  - 
\{\Phi_k^{(0)\dagger},\tilde \Phi_k^{(n)}\} V^{(n)} \right)
+{\rm h.c.}
\right ]_D
&\nn&
+ [W]_F  + [W^*]_F , 
\eea
where  the superpotential is:
\bea
&
W =  - i \sqrt{2} \sum_n \sum_{ijk} \epsilon_{ijk} {\rm Tr } \left [
4 g_0 v_k \sin \left({ n \pi \ta_k \over \Gamma}\right) \Phi_i^{(n)} \tilde \Phi_j^{(n)} \right. &\nn&   
- g  \cos \left({ n \pi \ta_k \over \Gamma}\right)
 \left ( [\Phi_k^{(0)},\Phi_i^{(n)}] \Phi_j^{(n)}  +  
[\Phi_k^{(0)},\tilde \Phi_i^{(n)}] \tilde \Phi_j^{(n)} \right)
&\nn& \left .
+  g  \sin \left({ n \pi \ta_k \over \Gamma}\right)
 \left ( \{ \Phi_k^{(0)},\Phi_i^{(n)} \} \tilde \Phi_j^{(n)}  - 
\{\Phi_k^{(0)},\tilde \Phi_i^{(n)}\} \Phi_j^{(n)} \right)  \right ].
\eea

 One immediate consequence of the custodial supersymmetry in the lagrangian  is that universal vevs are a flat direction at one-loop. Indeed, for $\langle \phi_{i,p} \rangle = v_i {\bf I}$, ${\rm STr} M^{2q} = 0$ and  the one-loop effective potential is zero  for  all such configurations. 
  The presence of the custodial supersymmetry is also sufficient to ensure the vanishing  of  one-loop corrections to the zero-mode masses. A mass-splitting of the zero-mode multiplets can appear only at the two-loop level and we expect the supersymmetry breaking scale to be suppressed $M_{\rm SUSY} \ll v \ll \Lambda$.   Supersymmetry is explicitly violated by triple and quartic self-interactions of the heavy modes. Thus, the heavy modes masses are not protected against logarithimically divergent corrections (earlier in the paper we have shown that quadratic divergences are absent at one-loop).  
\vskip 1cm
Let us now comment on the stringy interpretation of the results obtained 
within the field theoretical framework. 
The daughter theory is the low-energy field theory of branes located at the fixed point of
an  orbifold \cite{DOMO,GR,KASI,LANEVA,KLWI}. 
The low energy degrees of freedom on a brane are those combinations 
of the open string states that are invariant 
under the action of  $Z_\Gamma$. As discussed in 
\cite{KLWI} and \cite{ARCOKA}, for supersymmetric models interesting things happen when one 
moves  a stack of $n$  $D3$ branes at a distance $d$ away from the 
fixed point. Due to the $Z_\Gamma$ symmetry there are $\Gamma$ copies 
of the stack, spaced symmetrically in the transverse directions around the fixed point. It turns out that 
 moving the stacks of $n$ $D3$
branes from the origin is, from the field theory point of view, equivalent to going to the Higgs  branch of
the theory, where the scalars in off-diagonal subblocks with $\tilde a_i\ne 0$
acquire equal vacuum expectation values proportional to the $n \times n$ unit matrix, 
$\langle \phi_{i,p} \rangle =v \; {\rm \bf 1}_{n \times n}\;\; (v_i = v)$.
 Hence  moving the branes away from the fixed point leads to the  deconstruction phase of the field theory, with the gauge group broken down to its diagonal subgroup.  Since we discuss a field theory which is the low energy effective theory of such string models, this implies an extension 
of the results of \cite{ARCOKA} to nonsupersymmetric orbifoldings. 
Among various interesting points, it was shown in \cite{ARCOKA} that in the large $\Gamma$ limit, when the distances between images of the stack are much smaller than $d$, one can redefine the orbifold metric in such a way, 
that consecutive boson-fermion degenerate mass 
levels correspond to open strings winding around a circular direction of the transverse geometry.
 The geometric picture given in \cite{KLWI,ARCOKA} allows for the 
straightforward computation of the massive string spectrum:
\begin{equation}
m_n^2= 4 \frac{d^2}{l_s^4}\sum_{i=1}^3\sin^2( \frac{n \pi \tilde a_i}{\Gamma}) \end{equation}
where $l_s$ is the string scale and the shifts $\tilde a_i$ represent the action of $Z_\Gamma$ on the three complex coordinates. When all vevs are equal, this is precisely the field theoretical spectrum given by the formula (\ref{spectr}). 

 It is instructive to summarize the order in  which various scales appear when one increases the energy in the deconstruction phase of the field theoretical model. The first scale one encounters is the fictitious compactification scale\footnote{Strictly speaking we have here in mind the large $\Gamma$ limit. We put $a^2=\sum_i \tilde a_i^2$}
$1/R_5=agv/\Gamma$. At this scale a seeming  fifth dimension opens up and one sees 
the tower of Kaluza-Klein states with masses of order $1/R_5$. Hence above this scale
the theory looks five-dimensional.
Moreover the spectrum of massive states is determined by the custodial supersymmetry.
This picture holds up to the deconstruction scale $v$ where non-diagonal gauge bosons
become massless again. Above the deconstruction scale the theory 
is explicitly four-dimensional, nonsupersymmetric and renormalizable.
Quadratic divergences are absent at the one-loop level.
Also at one-loop the deconstruction scale is a flat direction of this four dimensional theory, hence it stays decoupled
from the string scale $1/l_s$. Moreover the compactification scale $1/R_5$ can be arbitrarily smaller
than the deconstruction scale and it is determined by the discrete parameter which is the 
order $\Gamma$ of the orbifold group $Z_{\Gamma}$.
\vskip 1cm

In this paper we have discussed the quiver theories which result from a 
nonsupersymmetric orbifolding of the ${\cal N}=4$ $U(K)$ gauge theories. 
We have shown, that in a generic situation these models, even though non-supersymmetric by construction, exhibit an improved UV behaviour - the quadratically divergent contributions to the effective potential vanish at the one-loop level. As the explicit calculation of the supertrace demonstrates, this is due to 
the fermion-boson cancellation that is inherited from the ${\cal N}=4$ mother theory. Thus, the hierarchy $v \ll \Lambda$ is protected at the one-loop level. One can realize in this way 
the solution of the hierarchy problem in its modest form, when the UV cut-off
is so low, say $100-1000$ TeV, that the one-loop absence of the quadratic divergences  becomes sufficient. 

When the gauge group  $U(n)^{\Gamma}$ of the daughter theory is  broken down to its diagonal subgroup we observe that parts of the lagrangian have the structure of a globally supersymmetric lagrangian. In consequence, at one-loop, universal vevs remain a flat direction and, also, the zero-mode multiplets do not obtain a mass splitting. Of course, the custodial supersymmetry is violated by interactions of the heavy modes and does not protect the UV behaviour of higher-loop contributions. Nevertheless, the class of theories we consider in this paper may be phenomenologically interesting, with the hierarchy of scales $M_{\rm SUSY} \ll v \ll \Lambda$. 
 
\vspace{1cm}
{\bf Acknowledgments:}
The authors thank A. Uranga, E. Dudas, J. Mourad and C. Savoy for fruitful discussions. 

The work of SP and AF was partially supported  by the EC Contract
HPRN-CT-2000-00148 for years 2000-2004. The work of ZL and AF was partially supported  by the EC Contract HPRN-CT-2000-00152 for years 2000-2004. The work of SP and ZL was partially supported  by the Polish State Committee for Scientific Research grant KBN 5 P03B 119 20 for years 2001-2002.


\end{document}